\documentclass[english,useAMS, usenatbib]{mn2e}
\usepackage[T1]{fontenc}
\usepackage[latin9]{inputenc}
\usepackage{units}
\usepackage{rotating}
\usepackage{url}
\usepackage{amsmath}
\usepackage{commath}
\usepackage{amssymb}
\usepackage{graphicx}
\usepackage{pdfpages}
\usepackage{esint}
\usepackage[authoryear]{natbib}
\usepackage{listings}

\makeatletter

\def\gtsima{$\; \buildrel > \over \sim \;$}
\def\ltsima{$\; \buildrel < \over \sim \;$}
\def\prosima{$\; \buildrel \propto \over \sim \;$}
\def\gsim{\lower.5ex\hbox{\gtsima}}
\def\lsim{\lower.5ex\hbox{\ltsima}}
\def\simgt{\lower.5ex\hbox{\gtsima}}
\def\simlt{\lower.5ex\hbox{\ltsima}}
\def\simpr{\lower.5ex\hbox{\prosima}}

\newcommand{\be}{\begin{eqnarray}}
\newcommand{\ee}{\end{eqnarray}}
\def\lsim{\,\lower2truept\hbox{${< \atop\hbox{\raise4truept\hbox{$\sim$}}}$}\,}
\def\gsim{\,\lower2truept\hbox{${> \atop\hbox{\raise4truept\hbox{$\sim$}}}$}\,}

\title[Galaxy formation  with GAMESH]{Galaxy formation with radiative and chemical feedback}
\author[L. Graziani et al.]{L. Graziani$^{1}$\thanks{E-mail: luca.graziani@oa-roma.inaf.it}, S. Salvadori$^{2}$, R. Schneider$^{1}$, D. Kawata$^{3}$, M. de Bennassuti$^{1}$, A. Maselli$^{4}$\\
$^{1}$ INAF Osservatorio Astronomico di Roma, Via Frascati 33, 00040, Monte Porzio Catone (RM), Italy\\
$^{2}$ Kapteyn Astronomical Institute, Landleven 12, NL-9747 AD Groningen, the Netherlands\\ 
$^{3}$ Mullard Space Science Laboratory, University College London, Holmbury St. Mary, Dorking, Surrey, RH5 6NT,UK \\
$^{4}$ EVENT Lab for Neuroscience and Technology, Universitat de Barcelona, Passeig de la Vall d'Hebron 171, 08035 Barcelona, Spain}

\begin{document}

\date{April 2014}

\pagerange{\pageref{firstpage}--\pageref{lastpage}} \pubyear{2009}

\maketitle

\label{firstpage}

\begin{abstract}

Here we introduce \texttt{GAMESH}, a novel pipeline which implements self-consistent radiative and chemical feedback 
in a computational model of galaxy formation. By combining the cosmological chemical-evolution model 
\texttt{GAMETE} with the radiative transfer code \texttt{CRASH}, \texttt{GAMESH} can post process realistic outputs 
of a N-body simulation describing the redshift evolution of the forming galaxy. 
After introducing the \texttt{GAMESH} implementation and its features, we apply the code to a low-resolution 
N-body simulation of the Milky Way formation and we investigate the combined effects of self-consistent radiative and 
chemical feedback. Many physical properties, which can be directly compared with observations in the Galaxy and its 
surrounding satellites, are predicted by the code along the merger-tree assembly.
The resulting redshift evolution of the Local Group star formation rates, reionisation and metal enrichment along with the 
predicted Metallicity Distribution Function of halo stars are critically compared with observations. We discuss the 
merits and limitations of the first release of \texttt{GAMESH}, also opening new directions to a full implementation 
of feedback processes in galaxy formation models by combining semi-analytic and numerical methods. 

\end{abstract}


\section{Introduction}

Since the pioneering work of \cite{1986MNRAS.221...53C}, the radiative feedback induced by cosmic 
reionisation has been recognised to have a strong impact on the formation and evolution of  
galaxies \citep{2000ApJ...542..535G, 2002MNRAS.333..156B, 2002MNRAS.333..177B, 2002ApJ...572L..23S, 
2006MNRAS.371..401H, 2008MNRAS.390..920O, 2009MNRAS.395L...6S, 2012ApJ...746..109L, 2014arXiv1406.6362S}, 
imprinting specific signatures in the observed properties of the Local Universe \citep{2012ApJ...753L..21B}. 

Semi-analytic models of galaxy formation (SAM) (see e.g. \citealt{2002MNRAS.333..177B,2002MNRAS.333..156B} or 
\citealt{2002ApJ...572L..23S}) and numerical simulations, both uni-dimensional \citep{1996ApJ...465..608T,
2000MNRAS.315L...1K, 2004ApJ...601..666D, 2013MNRAS.432L..51S} and three-dimensional (see \citealt{2000ApJ...542..535G, 
2006MNRAS.371..401H, 2008MNRAS.390..920O} and references therein) have been performed to determine 
the efficiency of the radiative feedback and the way it affects galactic star formation (SF) by either an 
extended ({\it Gradual}) or instant ({\it Sharp}) process in redshift; however a physically motivated prescription 
is still missing \citep{2014MNRAS.444..503N}. 
Models with a sharp SF suppression after reionisation generally fail to reproduce the present, observed luminosity 
function and predict an unacceptable gap between the faint and the bright populations, 
also showing that a redshift dependence in modelling feedback effects is necessary. 
The value of the critical mass below which galaxies should be strongly affected by photo-ionisation at fixed 
redshift is also subject to an intense debate: recent numerical studies \citep{2006MNRAS.371..401H, 2008MNRAS.390..920O} 
find about one order of magnitude lower masses compared to the original calculations \citep{2000ApJ...542..535G}, 
but all the conclusions critically depend on the reliability of the ionising background (UVB) adopted by the SAM. 

The effects of an UVB maintaining the Universe re-ionised can be also 
considerable on the structure and properties of the Local Group. The tendency of numerical 
simulations to over-predict the substructures on small scales compared to the observed population, 
the so called "missing satellite problem"  \citep{1999ApJ...524L..19M,1999ApJ...522...82K,
2012MNRAS.422.1203B}, is often interpreted as inability to reproduce the inefficient or suppressed 
star formation acting in the group of local dwarf galaxies, causing small halos to remain "dark", and then invisible to 
observations \citep{2009ApJ...703L.167A, 2010ApJ...710..408B,2014ApJ...794...20O}. 
Feedback processes have been invoked to justify this failure of the $\Lambda$CDM model, 
both mechanical \citep{2004ApJ...609..482K, 2009Natur.460..605D,2010ApJ...709.1138D} and 
radiative \citep{1992MNRAS.256P..43E,2000ApJ...542..535G, 2002MNRAS.333..177B, 2002ApJ...572L..23S,
2008ApJ...689L..41M, 2014MNRAS.444..503N, 2014MNRAS.440...50M}.   

A correct determination of the epoch of hydrogen reionisation on the scale of galaxy formation, 
self-consistently with the SF suppression operated by radiative feedback, could shed some light also 
on the nature of the ultra-faint dwarfs population found in the Sloan Digital Sky Survey (SDSS,  \citealt{2005AJ....129.2692W,2006ApJ...647L.111B,2006ApJ...643L.103Z,2006ApJ...650L..41Z}). 
The discovery of these galaxies, roughly doubling the number of known Milky Way (MW)
satellites, forced an update of the old models and opened new questions on whether this 
population is made by pre-reionisation fossils or it is just the lowest-luminosity tail of 
classical dwarfs. The answer will critically depend, again, on the way the gas ionisation is 
modelled and on how radiative effects are taken into account.

By using a data-calibrated model for the formation of the MW and its dwarf satellites,
which includes the presence of inefficient star-forming mini-halos and a heuristic prescription 
to account for radiative feedback, \cite{2009MNRAS.395L...6S} have been able to simultaneously 
reproduce the observed iron-luminosity relation and the Metallicity Distribution Function (MDF) 
of nearby dwarf galaxies, including the ultra-faint population. The authors argued that ultra-faint 
dwarf galaxies might represent the fossil relics of a once ubiquitous population of H$_2$-cooling
mini-halos that formed before reionisation ($z > 8.5$), similarly to what has been found by 
independent groups \citep{2008ApJ...689L..41M,2009ApJ...696.2179K, 2009ApJ...693.1859B,2009MNRAS.400.1593M, 2011ApJ...741...17B, 2011ApJ...741...18B} and in agreement with a series of more recent studies 
\citep{2012MNRAS.421L..29S, 2014MNRAS.437L..26S}. 
In particular, they find that a gradual suppression of the star-formation in 
increasingly massive (mini-)halos is necessary to match the observed iron-luminosity 
relation and the MDF of nearby dwarfs. Although this empirical relation is consistent with the 
minimum mass for SF settled by an advanced semi-analytic treatment of reionisation \citep{2014MNRAS.437L..26S}, 
the results still rely on the assumptions made on the nature and efficiency of feedback acting on SF halos. 
A more detailed modelling is certainly required to understand the {\it local} effects of in-homogeneous 
radiative feedback, and its impact on the luminosity function of MW satellites.

It should be noted, on the other hand, that the observed luminosity function of MW satellites is 
certainly not sufficient to fully constrain semi-analytic models as shown, for instance, by alternative 
approaches \citep{2010MNRAS.402.1995M,2010MNRAS.401.2036L}. The observed number and distribution of luminous 
satellites can be reproduced with accuracy also by assuming a sharp (i.e. instant) reionisation and a strongly 
mass-dependent star formation efficiency or by invoking other physical processes as supernovae feedback or tidal stripping.  

Recent investigations \citep{2009ApJ...703L.167A, 2010ApJ...710..408B, 2012ApJ...746..109L, 
2014ApJ...785..134L} try 
to make an advance in the treatment of the radiative feedback, mainly by relaxing the assumption 
of a uniform reionisation field \citep{2010MNRAS.402.1995M, 2010MNRAS.401.2036L}. These studies generally 
combine the merger-tree histories of advanced N-body simulations (e.g. \citealt{2008MNRAS.391.1685S}) with 
a semi-analytic radiative transfer (see for instance \citealt{2011MNRAS.414..727Z}) describing an 
in-homogeneous reionisation process. Despite the different approaches, the population of faint 
satellites is always highly sensitive to the adopted reionisation model.

We finally point out that the observed properties of the MW satellites in the Local 
Universe \citep{2009MNRAS.395L...6S,2014MNRAS.437L..26S} are the result of the global interplay 
between radiative, mechanical and eventually chemical feedback and a more comprehensive study is necessary to 
interpret the various observational signatures left during galaxy evolution: the origin and properties 
of the observed population of metal-poor stars in our galaxy and the low metallicity tail of 
the MDF.

In this paper we introduce \texttt{GAMESH}, a new pipeline integrating the latest release of 
cosmological radiative transfer code \texttt{CRASH} \citep{2013MNRAS.431..722G} with the semi-analytic 
model of galaxy formation \texttt{GAMETE}, powered by a N-body simulation (\citealt{2010MNRAS.401L...5S}, hereafter SF10). 
By following the star formation, metal enrichment and photo-ionisation in a self consistent way, 
\texttt{GAMESH} is an ideal tool to study the effects of photo-ionisation and 
heating in galaxy formation simulations and can address many of the questions discussed above. 
Into the bargain, the radiative transfer model adopted by \texttt{GAMESH} relies on a Monte Carlo scheme and
not only naturally accounts for the inhomogeneities of the ionisation process, but also allows a deep 
investigation of other radiative transfer effects, a self consistent calculation of the gas temperature,  
and a more accurate description of the stellar populations responsible for the Milky Way 
environment reionisation. 

In the present work the \texttt{GAMESH} pipeline will be applied to the Galaxy formation simulation 
introduced in \citet{2006ApJ...653..285S} and successfully post-processed by the semi-analytic code 
\texttt{GAMETE} (SF10), to derive the properties of the central galaxy and to compare them with the
available observations of the Milky Way system. Although 
the mass resolution of the adopted N-body does not provide a sufficient statistics of mini-halos to 
study the missing satellite problem, the robust set of results already 
obtained with the previous semi-analytic approach offers an excellent validation framework for the 
\texttt{GAMESH} pipeline and allows to appreciate the many advantages of the accurate radiative 
transfer treatment of \texttt{CRASH} coupled with the semi-analytic approach of \texttt{GAMETE}. 
A future work, already in place, will focus on high resolution simulations to address all the 
scientific problems discussed above with the adequate level of detail.

The paper is organised as follows. In Section 2 we describe the \texttt{GAMESH} pipeline by briefly 
introducing its components and by providing details on the feedback implementation. The Milky Way 
reionisation simulation is introduced in Section 3 and the results discussed in Section 4. Section 5 
finally summarises the conclusions of the paper.    

\begin{figure}
\includegraphics[width=0.50\textwidth]{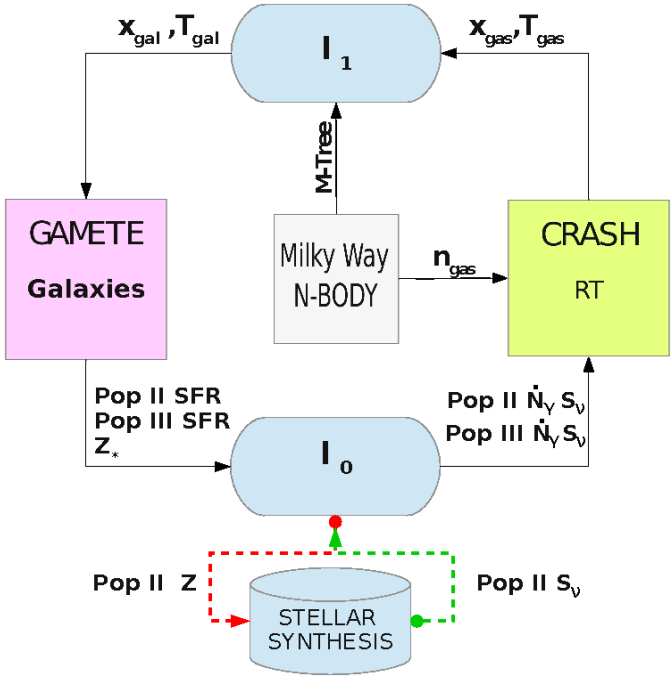}
\caption{\texttt{GAMESH} pipeline logic at fixed redshift $z_i$.
  The quantities $x_{\texttt{gal}}$, $T_{\texttt{gal}}$, $x_{\texttt{gas}}$ and $T_{\texttt{gas}}$ refer to the ionisation fractions 
  and temperatures of the gas in the grid cells containing galaxies and in the simulated domain, respectively.
  The gas number density projected into the grid used by \texttt{CRASH} is indicated as $n_{\texttt{gas}}$ 
  and the global set of information provided by the N-body merger tree as M-Tree.
  The quantities used by interactor $I_0$ are the star formation rates (SFR) and the metallicity of 
  the stars ($Z_{*}$), while the computed ionisation rates and spectral shapes per galaxy and population 
  are indicated as $\dot{N_{\gamma}}$ and $S_{\nu}$. 
  See text for more details.}
\label{fig:GAMESH-PIPELINE}
\end{figure}

\section{The GAMESH pipeline}

In this section we describe the work-flow of the \texttt{GAMESH} pipeline which integrates a N-body simulation, the semi-analytic code \texttt{GAMETE} (see Section 2.1 for more details), and the radiative transfer code \texttt{CRASH} (see Section 2.2); more details on each  pipeline module can be found in dedicated subsections.  

\texttt{GAMESH} models the galaxy formation process concatenating a series of snapshots provided by a N-body run at given redshifts $z_{i=0,...,N}$: the pipeline uses the physical quantities calculated at $z_i$ as initial conditions for the successive calculation at $z_{i+1}$. 
Along the redshift evolution, the feedback between star formation and RT is handled by two software modules called interactors $I_0$ (detailed in Section 2.3) and $I_1$ (see Section 2.4). 
$I_0$ transforms the galaxy star formation rates (SFR) predicted by \texttt{GAMETE}
into a list of ionising sources for \texttt{CRASH}, while $I_1$ uses the gas ionisation and temperature determined by the RT to establish a star formation prescription for the semi-analytic model implemented in \texttt{GAMETE}.
Hereafter we focus on the pipeline logic at fixed $z_i$ (see Figure 1); for an easier reading we also generically refer to the ionisation of the gas as $x_{\texttt{gas}}$, while a more specific notation will be adopted in Section 4 to discuss the ionisation fractions of hydrogen and helium. 

The initial conditions of the pipeline (ICs) are provided by the N-body simulation which assigns the simulation 
redshift $z_i$ to all the components, sets up the N-body merger-tree into $I_1$ and the gas number density $(n_ {\texttt{gas}})$ in the grid used by \texttt{CRASH} to map the physical domain. 

Once the ICs are set up, $I_1$ starts the simulation by creating a list of galaxies found in the merger-tree: each galaxy is identified by a unique ID and it associates the values of $x_{gal}$ and $T_{gal}$ found in the cell of the grid containing the galaxy center of mass (see section 2.4 for more details). This list is then processed by \texttt{GAMETE} to establish which galaxy can form stars, self-consistently with the metallicity, temperature and ionisation of the accreting gas.

As output of \texttt{GAMETE} we obtain a sub-sample of star forming galaxies, their SFR, stellar metallicity and population type. 
$I_0$ converts this sample into a list of \texttt{CRASH} sources by evaluating the galaxy positions on the grid,
their spectrum integrated ionisation rate $\dot{N_{\gamma}}$ and the spectral shape $S_{\nu}$. A stellar synthesis  database has been implemented in $I_0$ (see Section 2.3) to derive $\dot{N_{\gamma}}$ and $S_{\nu}$ from the stellar metallicity 
and the SFR. 

Once the properties of the radiating galaxies are established, the radiative transfer simulation starts propagating photons for a simulation duration 
corresponding to the Hubble time separating two snapshots, and it obtains the gas ionisation $x_{\texttt{gas}}$ and temperature $T_{\texttt{gas}}$ at redshift $z_i$. These quantities are finally used for the subsequent redshift $z_{i+1}$, by repeating the same algorithm.

Before moving to a more detailed description of the \texttt{GAMESH} components, we want to emphasise 
the advantages of our approach.
We first point out that the use of \texttt{CRASH} allows us to follow with accuracy the process of 
reionisation of the MW progenitors along the redshift evolution, by taking into account the intrinsic inhomogeneities
due to the gas clumps and by calculating the temperature history self-consistently. 
Moreover, radiative transfer effects, which generally lead to spectral hardening that may preferentially 
heat gas in over-dense regions, will be accounted for by the RT simulation without pre-assuming 
any propagation model. 

The adoption of \texttt{GAMETE} allows us to fully explore the interplay between reionisation, radiative feedback
and chemical evolution which can have very interesting and testable consequences (Schneider
et al. 2008). Using this approach, we can break the degeneracy between having 
gradual suppression and constant star formation efficiency or sharp suppression and mass-dependent 
star formation efficiency.

In the following sections more details on the pipeline components are provided. 

\subsection{GAMETE}

\texttt{GAMETE} \citep{2007MNRAS.381..647S} is a data-constrained semi-analytic model for
the formation of the Milky Way and its dwarf satellites \citep{2009MNRAS.395L...6S}.
The algorithm has been designed to study the properties of the first stars and
the early chemical enrichment of the Galaxy. The evolution of gas and stars inside 
each galactic halo of the merger-tree hierarchy (group of Dark Matter (DM) particles) is traced by assuming the 
following hypotheses:

\begin{itemize}

\item the gas is assumed of primordial composition at the highest redshift of the merger tree;
      
\item in each galaxy the SFR is proportional to the cold gas mass;

\item the contribution of radiative feedback is accounted for by adopting different prescriptions, depending on the problem at hand.  
       When the code runs in stand-alone mode (i.e. not in pipeline) \texttt{GAMETE} assumes instant reionisation (IReion, SF10), i.e. 
       stars form only in galaxies of mass $M_h > M_4(z)=3 \times 10^8 M_{\odot} (1+z)^{-3/2}$ ($M_h > M_{30}(z)=2.89 \times M_4(z)$) 
       prior to (after) reionisation (assumed complete at $z=6$). When \texttt{GAMETE} runs coupled with \texttt{CRASH}, 
       star formation is not regulated by the halo mass but by exact radiative feedback effects as predicted by the RT simulation 
       and detailed in Section 2.4.
       In this case the star formation efficiency in mini-halos (i.e. halos with $T_{vir} < 2 \times 10^{4}$K) is decreased $\propto T_{vir}^{-3}$
       as already detailed in \cite{2012MNRAS.421L..29S}, to mimic the effects of a LW background. 

\end{itemize}

Although recent versions of \texttt{GAMETE} \citep{2014MNRAS.437L..26S, 2014MNRAS.445.3039D} allow to follow the Pop\ III / Pop\ II transition by accounting for detailed chemical feedback (including the dust evolution) and exploring different Pop\ III IMFs and stellar lifetimes, as well as the effects of the in-homogeneous distribution of metals, the first release of \texttt{GAMESH} implements the simplified version of the chemical network as detailed below, for a better comparison with the results in SF10.

\begin{itemize}      
 
\item As assumed in the so called 'critical metallicity scenario' \citep{2002ApJ...571...30S, 2006MNRAS.369.1437S}
      low-mass stars form following a standard, Salpeter-like Initial Mass Function (IMF), when the metallicity of the gas 
      is $Z \ge Z_{cr}=10^{-4} Z_{\odot}$; when $Z < Z_{cr}$ we assume that Pop\ III stars
      form with a mass $m_{\texttt{Pop\ III}}=200 M_{\odot}$.

\item The enrichment of gas within the galaxies and in the diffused MW environment  
      (hereafter Milky Way Environment MWenv) is calculated by including a simple description of supernova (SN) feedback \citep{2008MNRAS.386..348S}. 
      By adopting the Instantaneous Recycling Approximation \citep{1980FCPh....5..287T} we assume that the gas 
      is instantaneously and homogeneously mixed with the atomic metals (see \citet{2007MNRAS.381..647S, 2010MNRAS.401L...5S} for 
      a discussion on the implications of this choice).
      
\end{itemize}

The interplay between N-body scheme and semi-analytic calculation, along the redshift evolution of 
the simulation, is detailed below. 
At each time-step we distribute the mass of gas, metals and stellar component of each halo among all 
the DM particles; this is used as ICs of the successive integration step. 
Metals ejected into the MWenv are treated in the same way so that the newly virialising halos have a chemical 
composition which depends on the level of enrichment the environment in which they are formed. To recover the spatial distribution of the 
long-living metal-poor stars at $z=0$, we also store their properties per DM particle.

\subsection{CRASH}

\texttt{CRASH} \citep{2001MNRAS.324..381C,2003MNRAS.345..379M,2005MNRAS.364.1429M,2009MNRAS.393..171M,2013MNRAS.431..722G} is 
a Monte Carlo based scheme implementing 3D ray tracing of ionising radiation 
through a gas composed by H and He and atomic metals (e.g. C, O, Si). 
The code propagates ionising packets with $N_{\gamma}$ photons per frequency  along rays crossing an arbitrary gas distribution 
mapped onto a Cartesian grid of $N_{c}^3$ cells. At each cell crossing,  
\texttt{CRASH} evaluates the optical depth $\tau$ along the casted path and the 
amount of absorbed photons $N_{abs} = N_{\gamma}(1-e^{\tau})$. $N_{abs}$ is used to 
calculate the ionisation fractions of the species $x_\texttt{{gas}} = x_\texttt{{i}} \in (x_{\rm HII}, 
x_{\rm HeII}, x_{\rm HeIII})$ and the gas temperature $T_{\texttt{gas}}$ of the crossed cell.  

The initial conditions of a \texttt{CRASH} simulation are assigned  
on a three-dimensional Cartesian grid and a cosmological box of linear 
size $L_{b}$, as detailed in the following list: 
\begin{itemize}
\item the number density of H ($n_{{\rm H}}$) and He ($n_{{\rm He}}$), the temperature of the gas
$T_{\texttt{gas}}$ and the gas ionisation fractions $x_\texttt{{i}}$ at the initial simulation time $t_{0}$\footnote{$n_ {\texttt{gas}}$ is calculated by projecting the particle distribution of the simulation
on a Cartesian grid of $N_{c}^3$ cells and by deriving the gas component via the 
universal baryon fraction.}; 
\item the number of point sources emitting ionising radiation ($N_{s}$), their position in
the Cartesian grid, their ionisation rate ($\dot{N_{\gamma}}$ in $\textrm{photons}$~$\textrm{s}^{-1}$)
and finally their spectral energy distribution (SED, $S_{\nu}$ in erg~s$^{-1}$~Hz$^{-1}$).  
The SED is assigned as an array of frequency bins providing the intensity of the radiation;
\item the duration of the simulation ($t_{f}$) and a pre-assigned set of simulation times
$t_{j}\in\left\{ t_{0},\ldots,t_{f}\right\} $ used to store the relevant physical quantities;
\item the intensity and SED of a UVB, if present. 
\end{itemize}

The interested reader can find more technical information about the latest implementation 
of \texttt{CRASH} and references on its variants in \cite{2013MNRAS.431..722G}.

\subsection{Radiation sources}

As explained above, \texttt{CRASH} requires the list of all the ionising sources 
present in the box grid. This is provided in the \texttt{GAMESH} framework by the 
\texttt{GAMETE}-to-\texttt{CRASH} module $I_0$, responsible of converting the properties of 
the star forming galaxies (Pop\ II/Pop\ III SFR, $Z_{*}$) into \texttt{CRASH} sources with 
specific spectral properties.

$I_0$ first maps the center of mass of each galaxy onto grid coordinates and then
converts the star formation rates into spectrum-integrated quantities $\dot{N_{\gamma}}$ 
depending on the stellar population. 

For Pop\ II stars we calculate ionisation rates and spectral shapes accordingly to \citet{1993ApJ...405..538B} 
and we assume a IMF in the mass range $[0.1-100]M_{\odot}$. 
A different spectral shape and ionisation rate is then associated with each of the Pop\ II star forming galaxy 
depending on its population lifetimes $t_{*}$ and stellar metallicity $Z_{*}$. The spectrum and ionisation rate 
is derived from a grid of pre-computed spectra integrated in specific lifetime bins   
$t_{*} \in \{0.001, 0.01, 0.1, 0.4, 1.0, 4.0, 13.0 \}$~Gyr and stellar metallicity $Z_{*} \in \{0.005, 0.2, 0.4, 1.0, 2.5\}Z_{\odot}$. 

For Pop\ III stars we assume an ionisation rate per solar mass, $\dot{N}_{\gamma}= 1.312 \cdot 10^{48}$ [$\textrm{photons}$~$\textrm{s}^{-1}/M_{\odot}$] \citep{2002A&A...382...28S}
corresponding to a stellar mass $M_{*} \sim 200 M_{\odot}$, averaged on a lifetime of about $2.2$~Myr. 
The SED of Pop\ III stars is simply assumed as black body spectrum at temperature $T_{BB} = 10^5$~K.

\subsection{Radiative feedback}

The radiative feedback on the star formation is implemented by the pipeline module $I_1$. 
This module is responsible for setting up the ionisation and temperature ($x_{\texttt{gal}}$, $T_{\texttt{gal}}$) of 
the gas in the galaxies, successively processed by \texttt{GAMETE} to establish their star formation. 

At the first redshift $z_0$ it is simply assumed that the medium is fully neutral (i.e. $x_{\texttt{gal}}=0$) 
and $T_{\texttt{gal}}=T_0(1+z_0)$, where $T_0$ is the value of the CMB temperature at $z=0$. 
During the redshift iterations $I_1$ computes  $x_{\texttt{gal}}$, $T_{\texttt{gal}}$ by first finding 
the cell containing the galaxy center of mass in the \texttt{CRASH} grid. As second step, it evaluates which of the 
surrounding cells best describe the environment from which the galaxy can get the gas to fuel star formation. 

The effective environment scale from which cold gas is fuelled into star forming galaxies is crucial to calculate 
the values of $x_{\texttt{gal}}$, $T_{\texttt{gal}}$ and to apply the radiative feedback. As in our pipeline the gas 
distribution surrounding halos relies on the spatial resolution of the RT (i.e. the cell size in \texttt{CRASH}: 
$\Delta L = L_{b}/N_{c}$), we compare the virial radius $R_{vir}$ of the galactic halo with $\Delta L /2$.  We simply assume
that if $2R_{vir} / \Delta L \leq 0.1$, the galactic environment is mainly set up 
in the cell containing its center of mass and we assign $x_{\texttt{gal}} = x_{\texttt{cell}}$, $T_{\texttt{gal}} = T_{\texttt{cell}}$. 
When, on the other hand,  $2R_{vir} / \Delta L > 0.1$ the gas reservoir of the galaxy could extend to the surrounding cells 
and then $T_{\texttt{gal}}$ is assigned to their volume averaged value; the value of $x_{\texttt{gal}}$ remains instead 
the one taken from the central cell.  It should be noted though, that the threshold value $0.1$ depends on the resolution of the
\texttt{CRASH} grid and must be tuned for increased resolutions grids, when necessary \footnote{Improving the accuracy of the 
galactic environment description would ideally require to selectively resolve the halo structures (i.e. a cell size $\Delta L << R_{vir}$), 
as obtained by using cell refinements around the galaxies. The spatial resolution adopted in the paper simulation is documented in Section 3.2}.
 
Once the temperature and ionisation fractions affecting the galactic star formation have been assigned, we first use $x_{\texttt{gal}}$ 
to evaluate the mean molecular weight $\mu$ of the gas and then to calculate the virial temperature of the galactic halo $T_{vir}$ (see the formulas in \citealt{2001PhR...349..125B}). 

Star formation in the galaxy is finally admitted in \texttt{GAMETE} if the condition $T_{\texttt{gal}} < T_{vir}$ applies, allowing 
accretion of cold gas in the galaxy to form stars. In the next version of \texttt{GAMESH} we will refine the star formation prescription also
accounting for gas cooling, metallicity dependence and dynamical time scales following the approach described in \citet{2014MNRAS.444..503N}; also see the Introduction for more references and details.

\section{Milky Way Reionisation simulation}

Here we describe the set-up of our Milky Way reionisation simulation. 
After a brief description of the N-body simulation and its parameters we 
detail the radiative transfer assumptions and the initial conditions.

Following \cite{2006ApJ...653..285S}, we adopt a cold dark matter 
($\Lambda$CDM) cosmological model with $h =0.71$, $\Omega_0h^2 = 0.135$,
$\Omega_{\Lambda}=1-\Omega_0$, $\Omega_bh^2 = 0.0224$, $n = 1$ and 
$\sigma_8 = 0.9$.

\subsection{The N-body simulation}

To study the Milky Way reionisation we adopt the snapshots and 
merger tree from a cosmological N-body simulation of the Milky 
Way-sized galaxy halo carried with \texttt{GCD+} \citep{2003MNRAS.340..908K}, 
which were used in \cite{2006ApJ...653..285S}. The detail of the 
simulation is described in \cite{2006ApJ...653..285S}; here we 
briefly summarise the simulation parameters necessary to understand 
the global \texttt{GAMESH} run.

The mass adopted for the N-body particles is 
$M_p=7.8\cdot10^5 M_{\odot}$ and the softening length used
is $540$ pc. The simulated system consists of 
about $10^6$ particles within the virial radius $R_{vir}= 239$ kpc 
for a total virial mass $M_{vir} = 7.7\cdot10^{11} M_{\odot}$. 
Note that the virial mass and virial radius estimated by observations 
for the Milky Way are $M_{vir} = 10^{12} M_{\odot}$ and $R_{vir} = 258$ kpc 
respectively \citep{2005MNRAS.364..433B}.

Using a multi-resolution technique \citep{2003MNRAS.346..135K}, the initial 
conditions at $z = 56$ are set up to resolve the region in a sphere within 
the radius of $R=4R_{vir}$ with the high-resolution N-body particles 
of mass $M_p=7.8\cdot10^5 M_{\odot}$ and with the softening length of 
$540$ pc; the other region is resolved instead with lower resolution particles. 

The simulation snapshots are provided with regular intervals every $22$ Myr 
between $z = 8$ and $17$ and every $110$ Myr for $z < 8$. The virialised DM 
halos are then identified by friend-of-friend group finder by assuming 
a linking parameter $b = 0.15$ and a threshold number of particles of 
$50$; the resulting minimum mass halo is $M_{h}=3.75 \times 10^7 M_{\odot}$. 

A validating N-body simulation at lower resolution, also including a star formation 
recipe, has also been performed to confirm that the assumed initial condition lead 
to a disc formation (see \citealt{2007ApJ...661...10B}). 

\subsection{The RT set-up}

The radiative transfer simulation is performed at each redshift $z_i$, 
on a box size of $2h^{-1}$ Mpc comoving, by emitting $N_p=10^6$ photon packets from 
each star-forming galaxy\footnote{Note that according to the pipeline algorithm each source could have a 'a priori' different spectrum assigned by $I_{0}$ as function of its lifetime and stellar metallicity.}. The box is mapped on a Cartesian grid of 
$N_c=128$ cells per cube side, providing a cell resolution of $\Delta L \sim 15.6h^{-1}$ kpc. 
The gas in the box is assumed of cosmological composition
(92\% H and 8\% He) and we neglect the metal ions and the effects of the metal cooling 
around the star-forming halos. We defer to a future application the extension of 
\texttt{GAMESH} to the metal ions. For a gas of cosmological composition ($n_\texttt{{HI}} \sim 92$\% $n_\texttt{{gas}}$) the ionising 
spectrum is usually assigned in the energy range $13.6$~eV$ \leq E_{\gamma} \leq 140$~eV. Note that the 
spectral database could be extended at any time to include harder spectra accounting for sources emitting  
higher energy photons. Note also that the \texttt{CRASH} calculation adopts the same 
cosmological parameters as in the N-body simulation to maintain consistency across 
the pipeline life-cycle.

The convergence of the Monte Carlo  sampling is guaranteed 
by an identical run with a factor of ten larger  $N_p$  and providing 
the same numbers within the fourth decimal of the volume average ionisation and 
temperature created in the box.         

\begin{figure}
\centering
\hskip -1.1truecm
\includegraphics[width=0.535\textwidth]{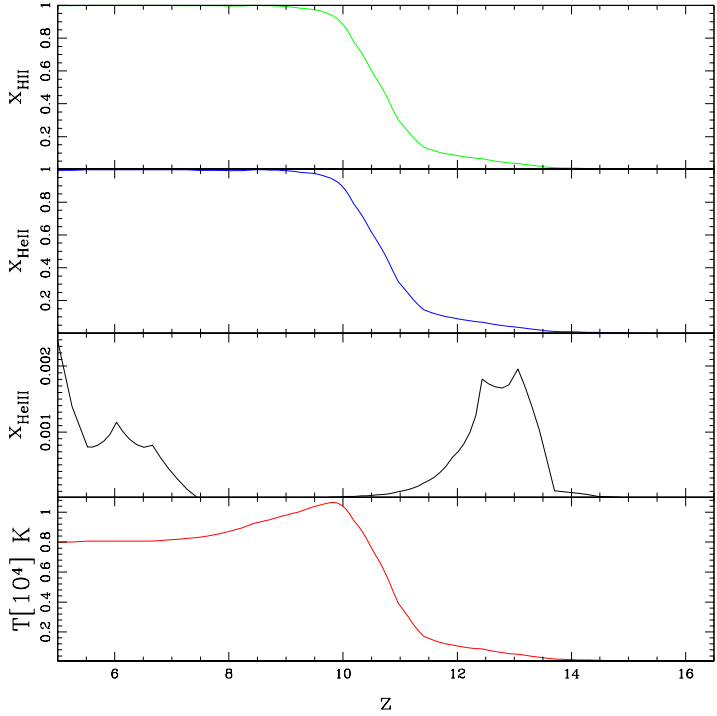}
\vspace{0.1truecm}
\caption{Redshift evolution of the volume averaged ionisation fractions and temperature 
         of the gas, down to $z \sim 5$. From top to bottom the values of $x_{\texttt{HII}}, 
         x_{\texttt{HeII}}, x_{\texttt{HeIII}}$, $T [10^{4}]$ K are shown in solid green, blue, 
         black and red lines respectively.}
\label{fig:RUN1-HISTORY}

\end{figure}

The simulation starts at $z_0 \sim 16$\footnote{This is the first redshift in which the applied FoF is able to identify the first halos. This redshift value depends on mass resolution of the simulation.} with uniform temperature $T_0 \sim 46$ K and  
we assume that the reionisation is completed when the volume-averaged ionisation 
fraction of the hydrogen reaches the value $x_{\rm HII} \geq 0.995$ against the gas 
recombination. Due to the large excursion in Hubble time, a gas recombination 
CASE-B is adopted in the simulation and the diffuse re-emission is neglected.

In the absence of a model for the UV background on the scale of the MW formation, we avoid
using an external cosmological UVB as done in other works present in the literature, 
and prefer to mimic a flux entering our box by applying periodic boundary conditions 
to the escaping radiation. While this choice guarantees that the photon mean free path is 
conserved when the box ionisation is advanced and the gas becomes transparent 
to the hydrogen-ionising radiation,  it should be noted that we do not account for an external 
ionising flux likely emitted by the background of QSO sources established below $z \sim 4$.

\section{Results}

In this section we discuss the results of our simulation, also comparing them with  
trends obtained adopting the assumption of instant reionisation at $z \sim 6$ as discussed 
in SF10. Hereafter, for an easier reading, we omit the label "$gas$" 
in the variable names referring to ionisation fractions and temperature of the gas and we 
comment on the single ionised species $x_\texttt{{i}} \in (x_{\rm HII}, 
x_{\rm HeII}, x_{\rm HeIII})$.

\subsection{Reionisation and temperature histories}

Here we investigate the evolution in redshift of the volume averaged ionisation fractions and temperature 
(in this section $x_{i}(z), T(z)$) resulting from the radiative transfer simulation. 

In Figure 2 we show the redshift evolution of $x_{\texttt{HII}}$ (top panel) and $x_{\texttt{HeII}}, x_{\texttt{HeIII}}$ (second and third panels from top) down to $z = 5$\footnote{Below this redshift a uniform ionised pattern of hydrogen is created and the values are less indicative. Note that we cannot discuss the helium reionisation due to the absence of a harder radiation background.}. 
$x_{\texttt{HII}}$ remains below $x_{\texttt{HII}} \sim 0.1$ when $z > 12$, afterwards it rapidly rises 
up to $x_{\texttt{HII}}\sim 0.5$ at $z\sim 11$ and reaches $x_{\texttt{HII}}\sim 0.9$ when $z\sim 10$.
The sudden increase in the hydrogen ionisation fraction in the redshift range $10 < z \leq 12$ can be ascribed both to 
the rising in the comoving SFR (Figure 4) and in the total number of emitting galaxies (Figure 5) but more likely to an 
advanced stage of overlapping of ionised regions, which makes the reionisation process intrinsically non-linear.

\begin{figure*}

\includegraphics[angle=0,width=1.08\textwidth]{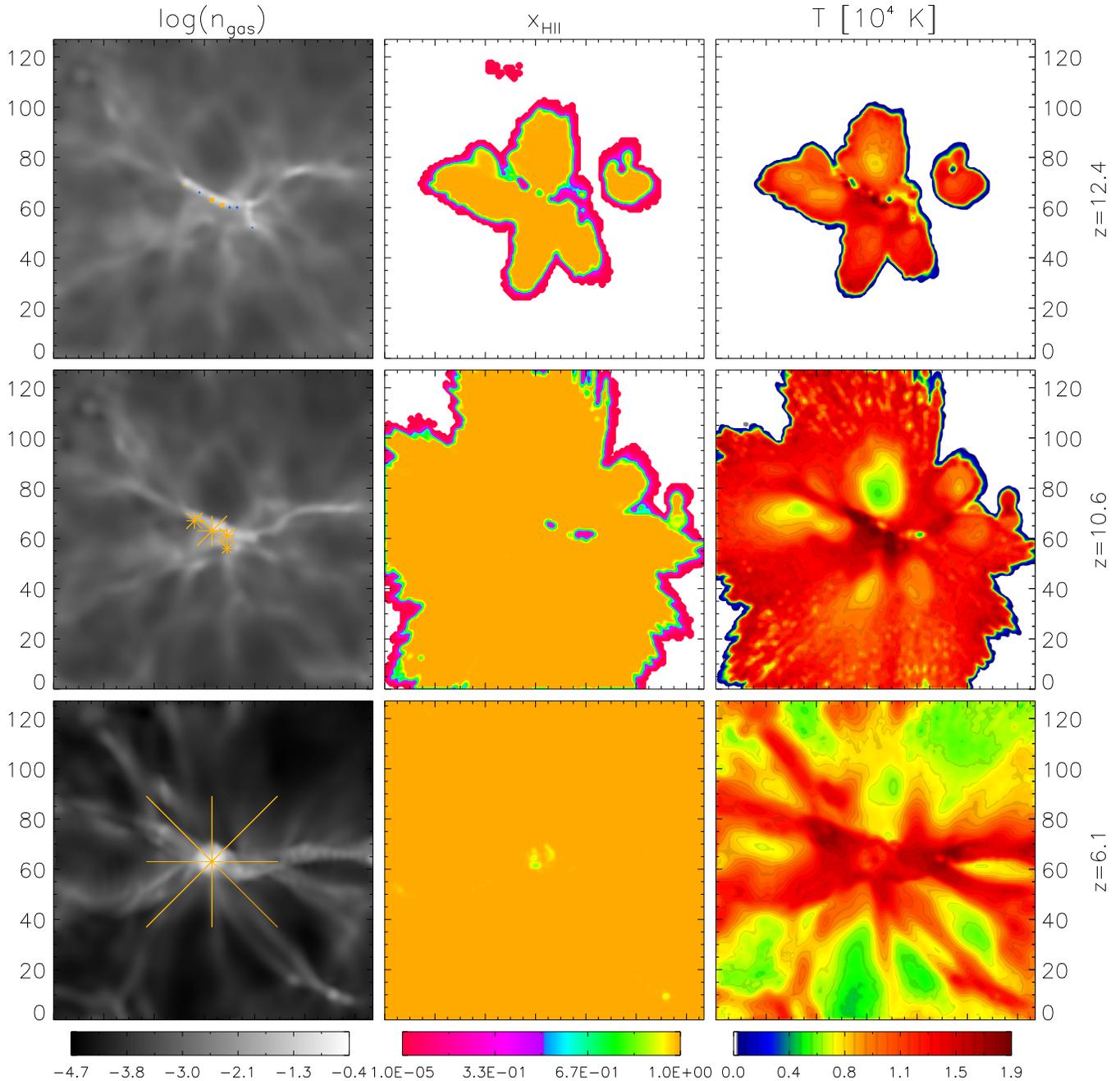}
\vspace{0.8truecm}
\caption{Slice cuts of $n_{gas}$ (first column), $x_{\texttt{HII}}$ (second column), $T$ (third column) at three 
different redshifts: $z \sim 12$ (top panels), $z \sim 11$ (middle panels), $z \sim 6$ (bottom panels). 
The value of the fields in the plane are represented as colour palettes and the distance units are expressed in 
number of grid cells along the x and y axis ($\sim 15.6h^{-1}$ kpc per cell). In the first-column panels star forming 
galaxies are symbolised as blue crosses when forming Pop\ III stars, while as gold-yellow asterisks for Pop\ II stars; the sizes of 
the symbols are scaled with the galaxy ionisation rate $\dot{N}_{\gamma}$}.

\label{fig:RUN1-HISTORY}

\end{figure*}

Below $z \sim 10 $ the hydrogen reionisation wades through and $x_{\texttt{HII}}$ increases up 
to $0.99$ by $z \sim 9$ and up to $0.999$ by $z \sim 5$. Even if not visible in the Figure, we 
point out that below $z \sim 8.5$ the value of the third decimal of $x_{\texttt{HII}}$ continues to oscillate in 
$0.997 \leq x_{\texttt{HII}} \leq 0.999$ due to the continuous unbalance in the gas ionisation, 
created by the ongoing gas collapse that enhances the recombination rate at the center, and by 
the in-homogeneous reionisation/recombination occurring in the volume. We consider the hydrogen 
reionisation completed at $z \sim 6.4$ when $x_{\texttt{HII}} > 0.995$, neglecting these fluctuations. 

The second and third panels from top, show the ionisation histories of helium. It is immediately 
evident that the He$\,{\rm {\scriptstyle I}}$ reionisation can be sustained only by the Pop\ II stars 
dominating below $z \sim 12$ while the reionisation of He$\,{\rm {\scriptstyle II}}$ never occurs 
at the considered Milky Way scale (here $2h^{-1} $ Mpc comoving) without the contribution of an external background of 
harder spectra (e.g. QSO). In fact, while the redshift trend of $x_{\texttt{HeII}}$ follows the one of
the hydrogen, $x_{\texttt{HeIII}}$ never reaches values higher than $x_{\texttt{HeIII}} = 0.002$ along 
the entire redshift range. As also pointed out for $x_{\texttt{HII}}$, the volume averaged ionisation fraction of 
He$\,{\rm {\scriptstyle II}}$ continues to oscillate in $0.97 < x_{\texttt{HeII}}< 0.99$ below $z \sim 8.5$,
confirming that a harder spectral component is required to sustain full helium ionisation against 
gas recombination. Note that a bump in the value of $x_{\texttt{HeIII}}$ (third panel from top) 
occurs in the redshift range $12 < z < 15$, tracing the presence of Pop\ III stars (see 
also Figure 4 ) with their harder spectral shapes. Note also that the volume averaged value of 
$x_{\texttt{HeIII}}$ remains always negligible over the entire simulation: after the transition from Pop\ III to Pop\ II 
stars the presence of full ionised helium is so closely confined to the sources that is not visible in the panel, 
except at z < 7, when hydrogen ionisation is complete and harder photons ($E_{\gamma} > 54.4$ eV)
are free to cross the box many times and to create a very small ionisation fraction of fully ionised helium.  

The redshift evolution of the volume-average temperature is shown in the bottom panel of Figure 2 in [$10^4$ K] scale. 
Before $z \sim 10$, the temperature rises from the initial value $T \sim 50$ K up to $T \sim 10^{4}$ K; 
note that the rapid evolution of Pop\ III stars in this redshift range is so small (not visible in the plot) that 
it cannot change the global trend of the average temperature. 
In the redshift interval $9 < z < 10$ the gas stabilises at a average photo-ionisation equilibrium temperature of 
$T \sim 10^{4}$ K, before its average value starts decreasing by some $10^3$ K during the successive evolution. 
Oscillations of the order of $10^3$ K are present below $z \sim 5$ (not shown in the plot), due to the competitive 
effects of gas clumping toward the center, adiabatic cooling due to the cosmological expansion, and finally 
in-homogeneous distribution and properties of the ionising galaxies. Even if 
not visible in this plot we report a final average temperature of  $T = 7.9 \times 10^3$ K at $z=0$ but we remind 
that this value is certainly underestimated in our simulation due to the absence of high energy photons from the QSO 
background established on the large scale below $z \sim 4$.

To visually illustrate how the reionisation process evolves in space and redshift, we show in Figure 3 three slice cuts of our volume (of $2h^{-1}$ comoving Mpc side length) taken at redshifts $z \sim 12, 11, 6$ (see panels from top to bottom). Panels in the first column show $log(n_{gas}(z))$ as gray gradient from black ($n_{gas} \sim 2 \times 10^{-5}$ cm$^{-3}$) to white ($n_{gas} \sim 0.4$ cm$^{-3}$); in the second column $x_{\texttt{HII}}(z)$ is shown by a colour palette from violet ($x_{\texttt{HII}} = 10^{-5}$) to orange ($x_{\texttt{HII}} = 1.0$), and finally the pattern of $T(z)$ is drawn in the third column with a different colour palette from white ($T \sim 50$ K) to dark red ($T \sim 1.9\cdot10^{4}$ K). 
In all the panels, the distances are shown in cell units ($\sim 15.6h^{-1}$ kpc per cell) along the x and y directions. 

Superposed to the number density maps we also show the star forming galaxies (SFG) present in the surface of choice: galaxies forming Pop\ III stars are symbolised as blue crosses (top panel), while gold-yellow asterisks are associated with the presence of Pop\ II stars. Also note that the size of each symbol is scaled with its $\dot{N}_{\gamma}$ and the position of star forming galaxies perfectly correlates with the gas filaments, 
falling toward the center where the central galaxy is progressively forming along the redshift evolution.

As consequence of the central position of the ionising sources, the reionisation proceeds inside-out. At high redshift (top panels), the low density regions surrounding the center of the box (black regions, first column) are easily ionised by the sources (orange pattern, second column). At the center of the image note that high-density clumps (white regions) are able to suppress photo-ionisation with a high gas recombination rate (magenta patterns). The average gas temperature (third column) tends to stabilise around $T\sim 1.4 \times 10^{4}$ K in the regions involved by the  overlap of ionised bubbles (yellow and red), mainly driven by the contribution of Pop\ III stars. $T$ rapidly decreases down to $T\sim 10^{2}$ K at the border of the ionised region (blue patterns) not reached by stellar radiation. Consistently with the ionised pattern, few high-density structures remain self-shielded and settle at very low temperature at the center image. Note that the presence of helium in our simulation significantly changes the gas photo-heating enhancing the contrasts between over-dense regions and voids.
At $z\sim 11$ (middle panels) the reionisation process is already quite advanced and the ionising radiation involves a large fraction of the box, including the under-dense regions far away from the center (green and yellow patters). Voids are rapidly ionised and their temperature stabilises around $T \sim 6\cdot10^{3}$ K, well below the central value $T\sim 10^{4}$ K. Finally, at $z\sim 6$ the entire volume reaches photo-ionisation equilibrium and its average temperature stabilises around $T \sim 8.0\cdot10^{3}$ K but the spatial pattern continues to show large variations from the gas filaments (branching off the central MW-type galaxy) and from the under-dense regions. 

Note that at the final redshift ($z \sim 6.1$) the central galaxy is seen as single source with highest ionisation rate of about $\dot{N}_{\gamma} = 2 \cdot 10^{54}$ photons/s and SFR $\sim 13 M_{\odot}$/yr (the total SFR of the emitting objects is $\sim 20 M_{\odot}$/yr ). In the other galaxies laying on the plane, the star formation is suppressed by radiative feedback (see Section 4.2.1 for more details)

As final comment, we point out that our simulation lacks the ionising contribution of an external UV background, especially at the helium ionising frequencies. While periodic boundary conditions are 
adopted to preserve the photon mean free path at all frequencies, the external UVB may play a relevant role because of the small size of the simulation box and its coarse mass resolution. 
On the other hand, the inclusion of a UVB is neither trivial nor straightforward.
First, the available UVB models (see for example \citealt{2012ApJ...746..125H} and references therein) are computed and calibrated on the large scale structure. Therefore, on the scales
of MW formation, both the intensity and the spectral shape of the UVB may be modified by RT effects. By means of the radiation tracking features of \texttt{CRASH3}, we will investigate these
effects in a future study. Here we only point out that an external UVB could only impact hydrogen ionisation above $z \sim 11$. In fact, Fig. 2 shows that below this redshift
the internal flux is sufficient to sustain hydrogen ionisation.

\subsection{Feedback on star formation}

\begin{figure}
\centering
\includegraphics[angle=0,width=0.50\textwidth]{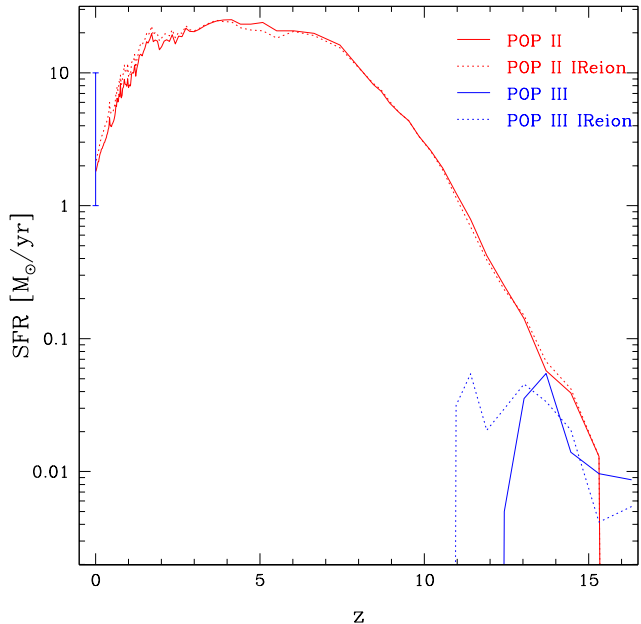}

\caption{Star formation rate for Pop\ II (solid red lines) and Pop\ III (solid blue) as function of $z$. As reference the values obtained with the instant reionisation approximation (IReion) are shown dotted lines and same colours for the two populations. The value of the SFR estimated by observation at $z=0$ is reported as error bar. See text for more details.}

\label{fig:SFR}

\end{figure}

The redshift evolution of the comoving SFR is shown in Figure 4, where the red lines refer to the contribution from
Pop\ II stars (highest curve) and the blue lines to Pop\ III. Dotted lines, with same colours, show the values obtained by adopting 
the instantaneous reionisation prescription (See Section 2.1).  
In the redshift interval $12.4 < z \lesssim 16$ the total SFR of our simulation (solid lines) increases up to $1.7 M_{\odot}$/yr thanks 
to both Pop\ III and Pop\ II stars, but note that at the highest redshift $z \sim 16.5$ (when the gas is mostly primordial) only population III 
objects contribute to the total SFR. 
Feedback from Pop\ III forming galaxies, both radiative and chemical, acts efficiently on the surrounding medium: their hard spectra locally increase the MWenv temperature inhibiting star formation in unpolluted clouds, while mechanical feedback rapidly pollutes the MWenv up to the critical metallicity $Z_{cr} = 10^{-4} Z_{\odot}$. Both effects contribute to their sudden disappearance, 
triggering the formation of Pop\ II stars below $z \sim 12.4$. At lower redshift Pop\ II stars 
cause an increase of the total SFR up to a value of SFR $\sim 25 M_{\odot}$/yr at the peak redshift $z \sim 4$. The successive 
evolution proceeds with an irregular but progressive decrease down to a value of SFR $\sim1.8 M_{\odot}$/yr at $z=0$, in good agreement with the observed data 
\citep{2011AJ....142..197C}.

\begin{figure}
\includegraphics[angle=0,width=0.50\textwidth]{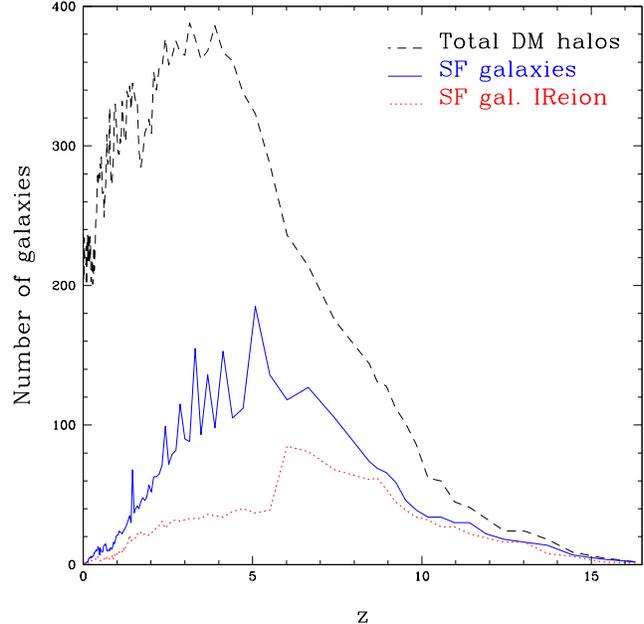}

\caption{Number of star forming galaxies as function of $z$ in the simulation (solid blue line). 
The dotted red line refers to the IReion case, while the dashed black line shows, as reference, 
the total number of DM halos found in the N-body simulation.}

\label{fig:SFHALOS}

\end{figure}

It is interesting to compare these results with the ones obtained with the instant reionisation (IReion) approximation for both populations. 
We remind that in IReion approximation \texttt{GAMETE} accounts for feedback by assuming that only Lyman-$\alpha$ halos having 
$M_{h} > M_{4}$ can form stars at $z > 6$, while at $z \sim 6$ the volume is suddenly ionised and a different threshold applies. 
For $z < 6$ only halos with circular velocity greater than $30$ kms$^{-1}$ (i.e. $M_{h} > M_{30}$) can trigger the star-formation (see SF10). 

By comparing the solid and dotted lines in Figure 4 we immediately conclude that the suppression of star formation operated by the IReion is
quite appropriated to reproduce the SFR of Pop\ II stars in a wide redshift range but it over-estimates the SFR at $z=0$ by 15\%. 
Note also that the stronger radiative feedback effects at high redshift in Ireion lower the SFR of Pop\ III stars and delay the transition between the two stellar populations.

\subsubsection{Feedback statistics}

The effects of the radiative feedback on the statistics of star forming galaxies are shown in 
Figure 5 as function of redshift $z$. The solid blue line indicates the number of SFG computed 
in our simulation, while the dotted red line refers to the IReion case. 
The total number of DM halos found in the N-body simulation in use is shown as dashed black line.

By comparing solid and dotted lines, we first note that at the current resolution of the simulation, 
which does not resolve the full range of masses that sample the mini-halos regime, the condition $M_{h} > M_{4}$ could 
appear quite appropriate to reproduce the statistics of the galaxies with suppressed star formation 
at very high redshifts ($z>10$). On the other hand, a careful comparison with Figure 4 immediately 
shows a different star formation rates of Pop\ III galaxies (compare blue lines in 
Figure 4).

This discrepancy can be ascribed to a combination of reasons which involve radiative, mechanical,
and chemical feedback. Also note that the IReion model cannot 
account for in-homogeneous feedback on high-redshift low mass objects (which are likely to host Pop\ III stars)  
but indiscriminately applies to {\it all} the galaxies with $M_{h} > M_{4}$ 
present in the volume. At high redshift this assumption is particularly incorrect because the SFG are still 
quite separated in space and then their radiative feedback acts very selectively, only affecting their surrounding neighbourhoods.
See for example how dis-homogeneous is the bubble overlap in the third panels of Figure 3 and how many 
regions of the simulated volume remain unaffected by photo-heating. 

Note that the current mass resolution of the N-body simulation does not allow to draw  robust conclusions 
on the radiative feedback effects at high-redshift. This will be the subject of a future dedicated study 
(de Bennassuti et al. 2015, in prep.).

At lower redshifts the efficiency of the radiative feedback increases with the progress of reionisation in both models.  
In the redshift interval $6 < z \lesssim 9$ the \texttt{GAMESH} run predicts that around 50\% of 
the galaxies have been affected by radiative feedback and this value increases up to 65\% by $z \sim 4$, 
with a jagged trend due to both oscillations in the number of potential galaxies (dashed black line) and 
the in-homogeneous heating in space induced by the RT\footnote{As commented in the previous paragraph, only with a higher mass and spatial resolution simulation we can fully understand the impact of an in-homogeneous reionisation process on the statistics of SF galaxies}.
Below $z=4$ the suppression of star formation in small galaxies progresses rapidly allowing few remaining 
objects in the box to form stars, as observed in the Milky Way environment.

\begin{figure*}

\includegraphics[angle=0,width=1.08\textwidth]{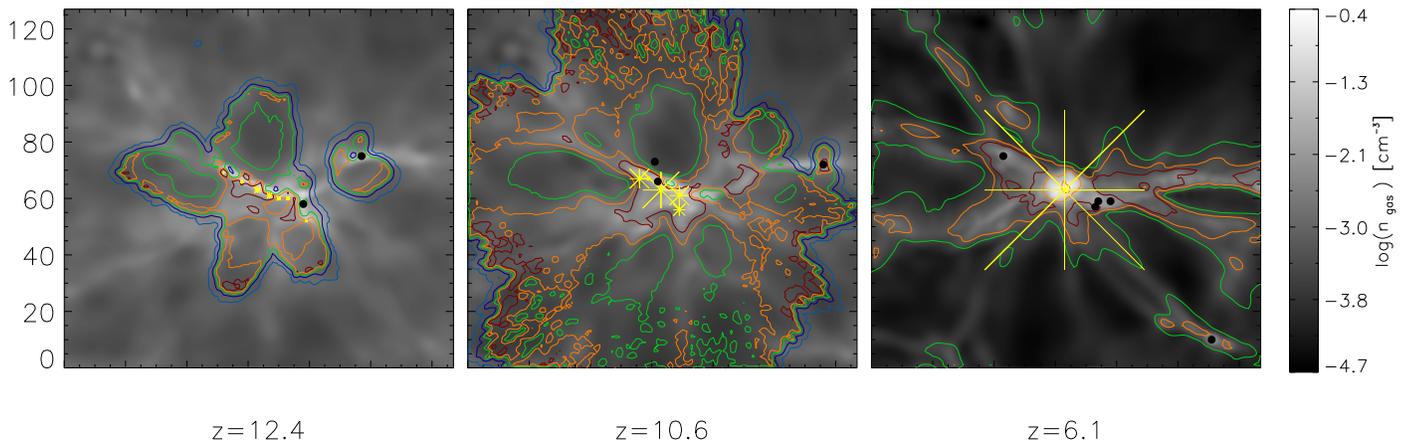}
\vspace{0.8truecm}
\caption{Slice cuts of the gas number density $n_{gas}$ with superposed temperature contour-plots:
$T \sim 100$ K (cyan), $T \sim 4 \times 10^{3}$ K (blue), $T \sim 10^{4}$ K (green), $T \sim 1.3 \times 10^{4}$ K (orange) 
and $T \sim 1.5 \times 10^{4}$ K in dark-red. 
All the star forming halos in the plane are represented by yellow asterisks, while black dots indicate halos in the 
plane in which star formation is suppressed by radiative feedback.}
\label{fig:RUN1-HISTORY}

\end{figure*}
 
Note the difference in the statistics of SFG below $z \sim 6$ obtained in the IReion approximation.
In fact a realistic radiative feedback operates with a more gradual suppression in time, due to the
in-homogeneous nature of the bubble overlap in space. In $3<z<6$ this difference induces an error 
in the number of SFG as high as 40\% between the two cases (compare solid blue line with dotted red line). 
On the other hand this statistical discrepancy between models is not reflected in both the total SFR (Figure 4) and 
the MWenv enrichment (see Figure 8) which are very similar in this redshift range, also indicating 
that the high-mass galaxies surviving the IReion prescription provide a dominant contribution to these quantities.

Figure 6 visually shows how the in-homogeneous radiative feedback operates in space, by comparing the same slice cuts and redshift sequence
commented in Figure 3. Here, on top of the gas number density field we have traced few solid line iso-contours of the gas temperature: 
$T \sim 100$ K in cyan, $T \sim 4 \times 10^{3}$ K in blue, $T \sim 10^{4}$ K in green, $T \sim 1.3 \times 10^{4}$ K as orange 
and finally $T \sim 1.5 \times 10^{4}$ K in dark-red. 
All the star forming galaxies are symbolised by yellow asterisks, while the
galaxies in which star formation has been suppressed by radiative feedback are shown as black filled circles.

By comparing the three panels it is immediately evident how the inhomogeneities in the reionisation process induce large differences 
in randomly suppressing star formation (left and middle panel) before a UVB is fully established (right panel). At high redshifts 
few suppressed SFGs (black circles) are trapped in regions with $T \sim 10^{4}$ K, not necessarily connected with the emitting galaxies. 
Note for example the galaxy on the right side of the middle panel: even far from the center it is surrounded by gas with high temperature, 
that likely originates from more complicated three-dimensional effects created by the RT. 
On the other hand at $z\approx 6$ (right panel)  the medium is already pervaded by a quasi-homogeneous UVB and then all the galaxies below a critical 
mass are easily suppressed in the entire domain. To study if a critical mass for the suppression of SFGs is statistically restored against 
the non-linearity induced by the RT, we should rely on a higher resolution simulation 
providing a better statistics for both Lyman-$\alpha$ and H$_{2}$-cooling halos, as well as better spatial resolution. With this 
study we will able to provide a numerically motivated recipe for semi-analytic models. We then defer 
this point to the next work (de Bennassuti et al. 2015, in prep.) and limit the discussion to this figure, as an illustrative example of how radiative feedback implemented 
in \texttt{GAMESH} works. 

\subsubsection{Local versus environmental feedback}

As introduced in Section 2.4, the radiative feedback defined in \texttt{GAMESH} depends on the spatial extension of the
galactic environment feeding cold gas for star formation. Depending on the virial radius of the dark matter halo hosting each galaxy and
the spatial resolution of the RT grid, the feedback could act {\it locally} to the sources (i.e. accounting just for the 
temperature in the galaxy cell ($2R_{vir} \leq 0.1 \Delta L$)) or it could act {\it globally} involving larger scales 
(i.e. cells surrounding the one containing the galaxy) when $2R_{vir} > 0.1 \Delta L$.

The statistics of these two mechanisms can help to understand the importance of 'local' versus 'global' feedback 
along the progress of reionisation, establishing a H-ionising uniform UVB. In Figure 7 we show the percentage of SFG (solid black 
line) together with the percentage of galaxies that are influenced by 'local' feedback (dashed red line) \footnote{These number are calculated with respect to the total number of halos found in the simulation.}. To understand which stellar population is affected by local or environmental 
effects we also show in blue dotted line the percentage of galaxies forming Pop\ II stars.   

\begin{figure}
\centering
\includegraphics[angle=0,width=0.50\textwidth]{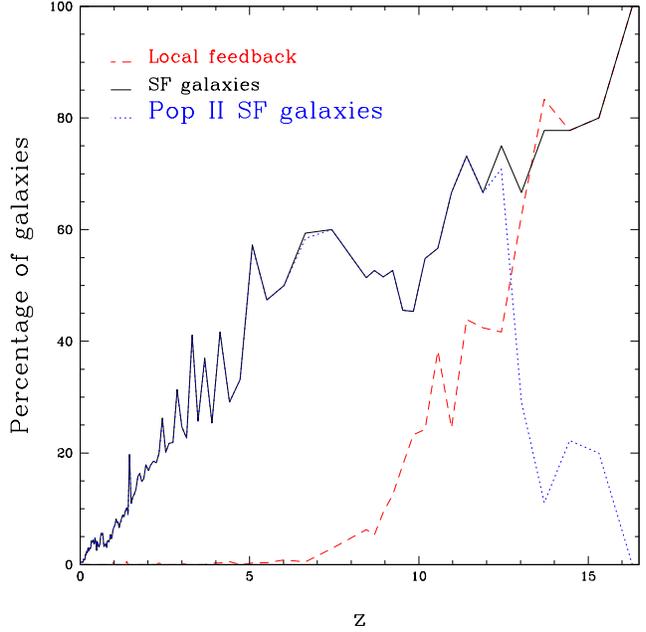}

\caption{Percentage of star forming galaxies as function of $z$ (solid black line) and percentage of galaxies
affected by radiative feedback in their own cell (dashed red line). The blue dotted line shows the percentage of galaxies 
forming mainly Pop\ II stars.}

\label{fig:SFR}

\end{figure}

As already commented in the previous figure, the percentage of SFG (solid black line) rapidly decreases from 
high to low redshifts following the progress of reionisation with a spiky and irregular trend down to 
$z\sim4$. The presence of a "plateau" in the redshift interval $6 < z < 9$ could mark the existence of an extended reionisation epoch in which 
the gas is kept in photo-ionisation equilibrium in the entire box at $T \sim 10^4$ K and the hydrogen ionisation 
fraction increases from $0.99$ up to $0.999$. The dashed line indicates that isolated, high redshift galaxies ($z>12.4$) are quite insensitive to their environment and the radiative feedback 
plays only a local role. A quick comparison with the blue dotted line clearly shows that these galaxies mainly 
form Pop\ III stars. After the population transition, Pop\ II forming galaxies are instead progressively affected 
by their environment: below $z \sim 8$, both the increase of the virial radii of the sources and a more uniform ionising 
background make the environment progressively more important and the feedback acts globally. At these redshifts the gas temperature is sustained by: i) the few satellite galaxies surviving radiative feedback, 
ii) the global UVB entering the box, and iii) the predominant emission from the central MW-type Galaxy. It is in fact clear 
from Figure 7 that less than 40\% of galaxies is capable to accrete gas from the surrounding reservoir to fuel the star 
formation process below $z \sim 4$.   

\subsection{Interplay with chemical feedback}

\begin{figure}
\centering
\includegraphics[angle=0,width=0.50\textwidth]{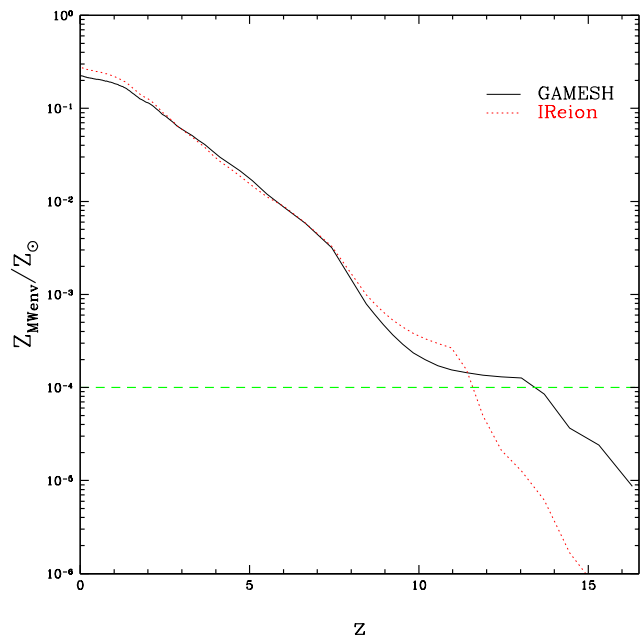}

\caption{Evolution in redshift of the Milky Way environment metallicity $Z_{MWenv}$ (solid black line) in solar 
metallicity units.  The values obtained by assuming instant reionisation are shown in dotted red line. The 
green dashed line shows, as reference, the value adopted for the critical metallicity to form Pop\ II stars.}

\label{fig:zIGM}
\end{figure}

In this section we briefly investigate how the reionisation process can leave specific signatures 
in the chemical evolution of the Milky Way environment and in the final properties of the MW at $z =0$ 
and, in particular, of the most metal-poor stars observed in the Galactic halo.

In Figure 8 we show the redshift evolution of the MWenv metallicity $Z_{MWenv}$\footnote{This value is defined as the metallicity of the medium surrounding the collapsed halos and it is calculated as ratio of the mass of metals over the total mass of the gas.} predicted by  
\texttt{GAMESH} (solid black line) and the corresponding values obtained by assuming instant 
reionisation (dotted red line). As already noted commenting Figure 4 and 5, the star-formation
history of Pop\ III stars is remarkably different when an accurate model for the radiative feedback is taken into 
account and their rapid burst, predicted by our simulation, is clearly imprinted in the metallicity 
of the diffuse, external medium. In fact, $Z_{MWenv}$ experiences a rapid increase at high redshift, 
reaching the critical value $Z_{cr}=10^{-4}Z_{\odot}$ already at $z \sim 14$. 

After a flat evolution corresponding to the Pop\ III to Pop\ II transition the increasing trend 
is restored by the increase in SFR of Pop\ II stars. 
As for the SFR (see Figure 4), $Z_{MWenv}$ becomes consistent with the value predicted by the 
IReion case in the redshift interval $3 < z \leq 8$ after which it flattens. Note that a 
15\%  difference in the SFR at $z=0$ due to an extended reionisation is reflected in 25\% decrease 
of the final metallicity $Z_{MWenv}$.

\begin{figure}
\centering
\includegraphics[angle=0,width=0.5\textwidth]{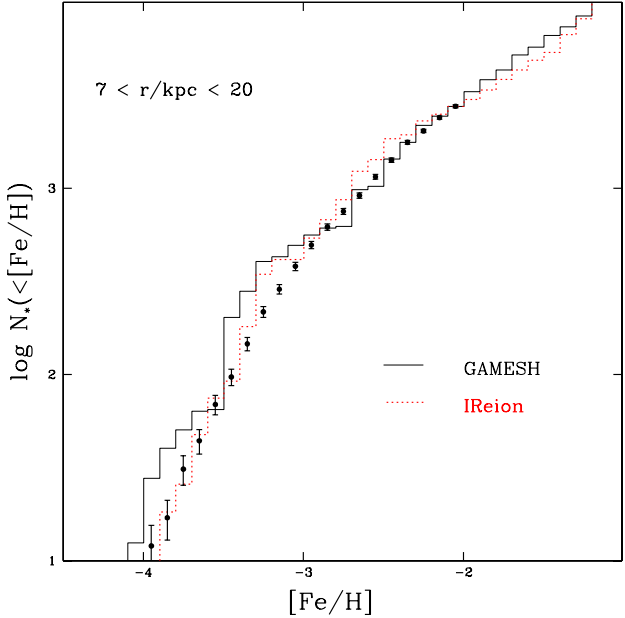}

\caption{MDF of the central galaxy calculated by \texttt{GAMESH} (solid black line) and in the 
IReion case (dotted red line) in the radial range $7 < r < 20$ kpc.  Black points indicate 
observed data taken from \citealt{2005ARA&A..43..531B}.}

\label{fig:MDF}
\end{figure}

It should be noted that the trend at high redshift is certainly un-physical because the metallicity of the medium 
surrounding the galaxies on our box scale is expected to increase more progressively in redshift 
from very low $Z_{MWenv}$ as correctly predicted by the IReion model. We thoroughly investigated what causes  
this trend, finding a concurrence of reasons. First, our N-body simulation 
does not provide a sufficient mass resolution to accurately predict the distribution of low 
mass mini-halos, which are expected to have played a relevant role in the early metal enrichment of the 
MW and its dwarf satellites (see for example \citealt{2009MNRAS.395L...6S, 2014MNRAS.439.2990S, 2014MNRAS.437L..26S}). 
As a consequence \texttt{GAMESH} tends to over-estimates the production/dispersion of metals in the first
star-forming objects. Second, as indicated by pure semi-analytic calculations that simultaneously reproduce 
different data-sets at $z=0$ (see \citet{2012MNRAS.421L..29S, 2014MNRAS.437L..26S}),
a better description of both the stellar life-times and the in-homogeneous dispersion of metals into 
the MWenv is required when the effect of an in-homogeneous reionisation are considered. Finally, on the 
radiative feedback side, the extension of the frequency range to a Lyman-Werner frequency band is necessary to 
correctly reproduce the star formation in H$_2$-cooling halos. 

The Metallicity Distribution Function (MDF) of ancient stars in the final MW-like galaxy, 
clearly reflects the different evolution of both the total SFR and $Z_{MWenv}$ at $z >8-10$,  
where the majority of [Fe/H] < -3 stars are formed \citep{2014MNRAS.445.3039D}. In Figure 9 
we compare the MDFs of $z=0$ stars in the radial range $1$kpc $< r < 20$ kpc for the IReion and \texttt{GAMESH} runs. Black points 
with errorbars show the available data for Galactic halo stars in the same Galactocentric ranges (see also SF10).

Both curves show a satisfactory agreement in the overall trend 
of the observed data, and the differences between the two runs are much smaller than the errors
induced when considering different MW merger histories \citep{2007MNRAS.381..647S, 2014MNRAS.445.3039D}.
While the \texttt{GAMESH} run fits the data for [Fe/H]$>-3$ (see \citealt{2005ARA&A..43..531B}), it clearly overproduces the number 
of low-metallicity stars, [Fe/H]$<-3$. This is due to the slower progress of metal enrichment, as reflected by the shallow increase of $Z_{MWenv}$ in the redshift range $8 < z < 13$ (see the plateau in Figure 8).  
We interpret this issue as a clear indication that the current resolution of the simulation does not 
allow to accurately trace star formation and radiative feedback at the highest redshifts.

\section{Conclusions}

In this work we investigated the early formation process of a MW-like galaxy by using \texttt{GAMESH}, a novel tool
combining for the first time in the literature, a N-body simulation with a detailed chemical and 
radiative feedback in a self-consistent framework. Many aspects of the formation process have 
been discussed along the cosmic time as the evolution of the SFR, the Pop\ III to 
Pop\ II transition, the statistics of progenitor 
halos in which star formation is suppressed, and finally the interplay between chemical and radiative feedback in 
setting up the chemical properties of the Local Group and the final MDF of the Galactic halo.

Even with a low resolution N-body simulation, which lacks in statistics and does not allow to draw
definite conclusions on many aspects of the problem, \texttt{GAMESH} has been proven to be
an ideal tool to provide both a comprehensive vision of the early formation of the Galaxy and its 
reionisation process, and to include an accurate self-consistent treatment of chemical, mechanical 
and radiative feedback in numerical simulations.

The capability of \texttt{GAMESH} to switch between pure semi-analytic treatment and inclusion 
of accurate RT enabled us to carefully compare and contrast all the results obtained by both models 
and to compare with observed properties of the Galaxy and the surrounding environment at $z=0$. In addition, the 
\texttt{GAMESH} algorithm is very general and does not limit its field of applicability to a Milky Way formation process; it can 
be extended and adapted to a wide range of galaxy and star formation related topics.

A future work, based on high mass and spatial resolution simulation, is certainly needed to 
establish with accuracy the epoch and extension of reionisation and its interplay with 
chemical feedback, the role of mini-halos and finally the signatures of global feedback processes 
on the stellar halo MDF and luminosity function of MW satellites. 

\section*{Acknowledgments}
The authors would like to thank the anonymous referee for his very constructive
comments. 
L.G. thanks R. Valiante for the invaluable support during the GAMETE refactoring and 
re-engineering. 
S. Salvadori acknowledges support from the Netherlands Organisation 
for Scientific Research (NWO), VENI grant 639.041.233. 
R.S., A.M. and S.S. wish to thank the Osservatorio 
Astrofisico di Arcetri where the project was originally conceived and developed in 
successive meetings.

The authors acknowledge Andrea Ferrara and Benedetta Ciardi for their very constructive comments. 
We also thank the 4C Institute at the Scuola Normale Superiore of Pisa for
the computational resources necessary to develop and test GAMESH. We also acknowledge PRACE 
\footnote{http://www.prace-ri.eu/} for awarding us access to the CEA HPC facility "CURIE@GENCI"
\footnote{http://www-hpc.cea.fr/en/complexe/tgcc-curie.htm} with the Type B project: High 
Performance release of the GAMESH pipeline.

The research leading to these results has received funding from the European 
Research Council under the European Union's Seventh Framework Programme 
(FP/2007-2013) / ERC Grant Agreement n. 306476.

\bibliographystyle{mn2e}
\bibliography{GAMESH}

\label{lastpage}
\end{document}